\documentclass[10pt,conference]{IEEEtran}

\IEEEoverridecommandlockouts

\usepackage{amsmath,amssymb,amsfonts}
\usepackage{algorithmic}
\usepackage{graphicx}
\usepackage{textcomp}
\usepackage[table]{xcolor}
\def\BibTeX{{\rm B\kern-.05em{\sc i\kern-.025em b}\kern-.08em
    T\kern-.1667em\lower.7ex\hbox{E}\kern-.125emX}}

\usepackage[colorinlistoftodos]{todonotes}
\usepackage{xspace}
\usepackage{threeparttable}
\usepackage{tabularx}
\usepackage{multirow}
\usepackage{booktabs}
\usepackage{graphicx}
\usepackage{arydshln}
\usepackage[hyphens]{url}
\usepackage{comment}

\newcommand{\sysname}{D2A\xspace}

\newcommand{\openssl}{\texttt{OpenSSL}\xspace}
\newcommand{\ffmpeg}{\texttt{FFmpeg}\xspace}
\newcommand{\nginx}{\texttt{NGINX}\xspace}
\newcommand{\httpd}{\texttt{httpd}\xspace}
\newcommand{\libav}{\texttt{libav}\xspace}
\newcommand{\libtiff}{\texttt{libtiff}\xspace}
\newcommand{\combined}{\texttt{combined}\xspace}

\newcommand{\infer}{Infer\xspace}

\usepackage{array, makecell}
\usepackage{bbding}
\usepackage{pifont}

\definecolor{darkpastelblue}{rgb}{0.47, 0.62, 0.8}
\definecolor{darkpastelgreen}{rgb}{0.01, 0.75, 0.24}
\definecolor{darkpastelred}{rgb}{0.76, 0.23, 0.13}
\definecolor{coolgrey}{rgb}{0.55, 0.57, 0.67}
\newcommand{\cmark}{\color{darkpastelgreen} \text{\ding{52}}}
\newcommand{\xmark}{\color{darkpastelred} \ding{56}}

\newcommand{\cellgreen}{\cellcolor{darkpastelgreen!10}}
\newcommand{\cellblue}{\cellcolor{darkpastelblue!15}}
\newcommand{\cellgrey}{\cellcolor{coolgrey!10}}

\usepackage{enumitem}

\begin{document}

\title{\sysname: A Dataset Built for AI-Based Vulnerability Detection Methods Using Differential Analysis}

\author{
\IEEEauthorblockN{Yunhui Zheng, Saurabh Pujar, Burn Lewis, Luca Buratti, Edward Epstein, Bo Yang, \\Jim Laredo, Alessandro Morari, Zhong Su}
\IEEEauthorblockA{{\bf IBM Research}\\
\{zhengyu, burn, eae, laredoj, amorari\}@us.ibm.com, \{saurabh.pujar, luca.buratti1\}@ibm.com, \\ \{yangbbo, suzhong\}@cn.ibm.com}
}

\maketitle

\begin{abstract}
Static analysis tools are widely used for vulnerability detection as they understand programs with complex behavior and millions of lines of code. Despite their popularity, static analysis tools are known to generate an excess of false positives. The recent ability of Machine Learning models to understand programming languages opens new possibilities when applied to static analysis. However, existing datasets to train models for vulnerability identification suffer from multiple limitations such as limited bug context, limited size, and synthetic and unrealistic source code. We propose D2A, a differential analysis based approach to label issues reported by static analysis tools. The D2A dataset is built by analyzing version pairs from multiple open source projects. From each project, we select bug fixing commits and we run static analysis on the versions before and after such commits. If some issues detected in a before-commit version disappear in the corresponding after-commit version, they are very likely to be real bugs that got fixed by the commit. We use D2A to generate a large labeled dataset to train models for vulnerability identification. We show that the dataset can be used to build a classifier to identify possible false alarms among the issues reported by static analysis, hence helping developers prioritize and investigate potential true positives first.
\end{abstract}


\section{Introduction}
\label{sec:intro}


The complexity and scale of modern software programs often lead to overlooked programming errors and security vulnerabilities. Research has shown that developers spend more than 50\% of their time detecting and fixing bugs \cite{LaToza_ICSE06, Murphy-Hill_TSE_15}. In practice, they usually rely on automated program analysis or testing tools to audit the code and look for security vulnerabilities. Among them, static program analysis techniques have been widely used because they can understand nontrivial program behaviors, scale to millions of lines of code, and detect subtle bugs \cite{SA4Bug, Ayewah_OOPSLA07, Yamaguchi_SP15, Smoke}. Although static analysis has limited capacity to identify bug-triggering inputs, it can achieve better coverage and discover bugs that are missed by dynamic analysis and testing tools. In fact, static analysis can provide useful feedback and has been proven to be effective in improving software quality \cite{Livshits_05, Guarnieri_ISSTA11}.

Besides these classic usage scenarios, driven by the needs of recent AI research on source code understanding and vulnerability detection tasks~\cite{Ulas_ICSM13, ALETHEIA, Koc_MAPL17, draper, vuldeepecker, sbabi, devign, Luca_bert, Suneja_icst2020, codeDiffForApiUsage}, static analysis techniques have also been used to generate labeled datasets for model training \cite{draper}. As programs exhibit diverse and complex behaviors, training models for vulnerability detection requires large labeled datasets of buggy and non-buggy code examples. This is especially critical for advanced neural network models such as CNN, RNN, GNN, etc. 
However, existing datasets for vulnerability detection suffer from the following limitations:
\begin{itemize}[leftmargin=*,nolistsep, noitemsep]
    \item Almost all datasets are on a function level and do not provide context information (e.g., traces) explaining how a bug may happen. Besides, they usually do not specify the bug types and locations. In many cases, the function-level example does not even include the bug root cause.
    
    \item Some datasets (e.g. CGD~\cite{vuldeepecker}) are derived from confirmed bugs in NVD~\cite{NVD} or CVE~\cite{CVE}. Although they have high-quality labels, the number of such samples is limited and may be insufficient for model training.
    
    \item Synthetic datasets such as Juliet \cite{juliet} and S-babi\cite{sbabi} can be large. However, they are generated based on a few predefined patterns and thus cannot represent the diverse behaviors observed in real-world programs.
    
    \item There are also labeling efforts based on commit messages or code diffs. Predicting code labels based on commit messages is known to produce low-quality labels~\cite{draper}. Code diff based methods \cite{devign} assume all functions in a bug-fixing commit are buggy, which may not be the case in reality. More importantly, these approaches have difficulty in identifying bug types, locations, and traces.
\end{itemize}

On the other hand, static analysis can reason beyond function boundaries.  It's automated and scales well enough to generate large datasets from programs in the wild. For example, Russell et. al \cite{draper} applied the Clang static analyzer, Cppcheck, and Flawfinder to generate a labeled data set of millions of functions to train deep learning models and learn features from source code. In some sense, it is the most promising labeling approach as it can additionally identify bug types and locations while using traces as context information.

Despite the popularity in these scenarios, static analysis tools are known to generate an excess of false alarms. One reason is the approximation heuristics used to reduce complexity and improve scalability. In particular, static analysis tries to model all possible execution behaviors and thus can suffer from the state-space blowup problem \cite{ALETHEIA}. To handle industry-scale programs, static analysis tools aggressively approximate the analysis and sacrifice the precision for better scalability and speed. For example, the path-sensitive analysis does not scale well on large programs, especially when modeling too many path states or reasoning about complex path constraints. Therefore, path insensitive analysis that ignores path conditions and assumes all paths are feasible is commonly used in practice, which obviously introduces false positives.

These \textit{false positives greatly hinder the utilization of static analysis tools} as it is counterproductive for developers to go through a long list of reported issues but only find a few true positives \cite{Johnson_icse13, Muske_13}. To suppress them, various methods have been proposed (as summarized in \cite{Muske_16}). Among them, machine learning based approaches \cite{Kremenek_03, Jung_05, Yuksel_13, Hanam_14, ALETHEIA, Koc_17, Zhang_MAPL_17, Reynolds_17, Raghothaman_18, Koc_19} focus on learning the patterns of false positives from examples. However, training such models requires good labeled datasets. Most existing works manually generate such datasets by reviewing the code and bug reports. In our experience, this review process is very labor-intensive and cannot scale. Therefore, the datasets are relatively small and may not cover the diverse behaviors observed in reality.

To address these challenges, in this paper, we propose \sysname, a differential analysis based approach to label issues reported by static analysis tools as ones that are {\it more likely to be true positives} and ones that are {\it more likely to be false positives}. Our goal is to generate a large labeled dataset that can be used for machine learning approaches for (1) static analyzer false positive reduction, and (2) code understanding and vulnerability detection tasks. 
We demonstrate how our dataset can be helpful for the false positive reduction task.

In particular, for projects with commit histories, we assume some commits are code changes that fix bugs. Instead of predicting  labels based on commit messages, we run static analysis on the versions before and after such commits. 
If some issues detected in a before-commit version disappear in the corresponding after-commit version, they are {\it very likely to be real bugs} that got fixed by the commit. If we analyze a large number of consecutive version pairs and aggregate the results, some issues found in a before-commit version never disappear in an after-commit version. We say they are {\it not very likely to be real bugs} because they were never fixed.
Then, we de-duplicate the issues found in all versions and adjust their classifications according to the commit history. Finally, we label the issues that are very likely to be real bugs as {\it positives} and the remaining ones as {\it negatives}.  We name this procedure {\it differential analysis} and the labeling mechanism {\it auto-labeler}. Please note that we say the reported issues are very likely to be TPs or FPs because the static analyzer may make mistakes or a bug-fixing commit was not included for analysis. We will discuss this in more detail in Sec.~\ref{sec:dataset}.

We run the differential analysis on thousands of selective consecutive version pairs from \openssl, \ffmpeg, \libav, \httpd, \nginx and \libtiff. Out of 349,373,753 issues reported by the static analyzer, after deduplication, we labeled 18,653 unique issues as positives and 1,276,970 unique issues as negatives. Given there is no ground truth, to validate the efficacy of the auto-labeler, we randomly selected and manually reviewed 57 examples. The result shows that \sysname improves the label accuracy from 7.8\% to 53\%. 

Although the \sysname dataset is mainly for machine learning based vulnerability detection methods, which usually require a large number of labeled samples, in this paper, we show it can be used to help developers prioritize static analysis issues that are more likely to be true positives. We will present AI-based code understanding approaches in another paper. In particular, inspired by \cite{ALETHEIA}, we defined features solely from static analysis outputs and trained a static analysis false positive reduction model. The result shows that we were able to significantly reduce false alarms, allowing developers to investigate issues that are less likely to be false positives first.
In summary, we make the following contributions:
\begin{itemize}[leftmargin=*, nolistsep, noitemsep]
    \item We propose a novel approach to label static analysis issues based on differential analysis and commit history heuristics.
    
    \item Given it can take several hours to analyze a single version pair (e.g. 12hrs for \ffmpeg), we parallelized the pipeline such that we can process thousands of version pairs simultaneously in a cluster, which makes \sysname a practical approach.
    
    \item We ran large-scale analyses on thousands of version pairs of real-world C/C++ programs, and created a labeled dataset of millions of samples with a hope that the dataset can be helpful to AI method on vulnerability detection tasks. 
    
    \item Unlike existing function-level datasets, we derive samples from inter-procedural analysis and preserve more details such as bug types, locations, traces, and analyzer outputs.  
    
    \item We demonstrated a use case of the \sysname dataset. We trained a static analysis false positive reduction model, which can effectively reduce false positive rate and help developers prioritize issues that are more likely to be real bugs.

    \item To facilitate future research, we make the \sysname dataset and its generation pipeline publicly available at \url{https://github.com/ibm/D2A}.

\end{itemize}



\section{Motivation}
\label{sec:moti}

In this section, we describe two use scenarios to show why building a good labeled dataset using static analysis can be useful for AI-based vulnerability detection methods.

\subsection{Existing Datasets for AI on Vulnerability Detection Task}
\label{sec:moti_dataset}

\begin{table*}[t]
 	\centering
	\caption{Publicly Available Datasets for AI on C/C++ Vulnerability Detection}
 	\vspace{-0.5em}
{\bgroup
\def\arraystretch{1}
\resizebox{0.95\textwidth}{!}{%
    \begin{tabular}{l c c c c c  c  c   c   c c  l } 
      \toprule
        {\bf Dataset} & 
        \makecell{ {\bf Example} \\ {\bf Type} } &
        \makecell{ {\bf Example} \\ {\bf Level} } &
        \makecell{ {\bf Whole Dataset} \\ {\bf Released} } & 
        \makecell{ {\bf Bug} \\ {\bf Type} } & 
        \makecell{ {\bf Bug} \\ {\bf Line} } & 
        \makecell{ {\bf Bug} \\ {\bf Trace} } & 
        \makecell{ {\bf Codebase} \\ {\bf Traceability} }& 
        \makecell{ {\bf Compilable} \\ {\bf Example} } &
        \makecell{ {\bf Generation} \\ {\bf Impl. Avail.} } &
        \makecell{ {\bf Labelling Method}} 
      \\ \midrule
            Juliet \cite{juliet}&
            synthetic &  
            function &   
            \cmark & 
            \cmark &       
            \cmark &
            \xmark &
            -- & 
            \cmark &
            -- &
            predefined pattern
        \\
            S-Babi \cite{sbabi}  &
            synthetic &  
            function &
            \cmark & 
            \cmark &       
            \cmark &
            \xmark &
            -- & 
            \cmark &
            \cmark &
            predefined pattern
        \\
            Choi et.al \cite{Choi_2017} &
            synthetic &  
            function &
            \cmark & 
            \cmark &       
            \cmark &
            \xmark &
            -- & 
            \cmark &
            \cmark &
            predefined pattern
        \\
            Draper \cite{draper} &
            mixed &
            function &
            \cmark & 
            \cmark &
            \xmark &
            \xmark &
            \xmark & 
            \xmark & 
            \xmark &
            static analysis 
        \\
            Devign \cite{devign} &
            real-world & 
            function &
            \xmark &
            \xmark &
            \xmark &
            \xmark &
            \xmark & 
            \xmark &
            \xmark &
            manual + commit code diff
        \\
            CDG \cite{vuldeepecker} &
            real-world & 
            slice &
            \cmark &
            \xmark &
            \xmark &
            \xmark &
            \xmark & 
            \xmark &
            \cmark &
            NVD + code diff
        \\
            \sysname &
            real-world &
            trace &
            \cmark &
            \cmark &
            \cmark &
            \cmark &
            \cmark &
            \cmark &
            \cmark &
            differential static analysis 
        \\
      \bottomrule
    \multicolumn{11}{l}{\textbf{\texttt{Note}}: To the best of our knowledge, there is no perfect dataset that is large enough and has 100\% correct labels for AI-based vulnerability detection tasks. Datasets} \\
    \multicolumn{11}{l}{\hspace{0.3in} generated from manual reviews have better quality labels in general. However, limited by their nature, they are usually not large enough for model training. } \\
    \multicolumn{11}{l}{\hspace{0.3in} On the other hand, the quality of the \sysname dataset is bounded by the capacity of static analysis. \sysname has better labels comparing to datasets labeled solely  } \\
    \multicolumn{11}{l}{\hspace{0.3in} by static analysis and complements existing high-quality datasets by the size. Please refer to Sec.~\ref{sec:moti_dataset} for details.} \\
    \end{tabular}
    }
\egroup}
\label{table:moti_dataset_comprison}
\vspace{-0.15in}
\end{table*}

Since programs can exhibit diverse behaviors, training machine learning models for code understanding and vulnerability detection requires large datasets. However, according to a recent survey~\cite{dl_vi_survey}, lacking good and real-world datasets has become a major barrier for this field. Many existing works created self-constructed datasets based on different criteria. However, only a few fully released their datasets.

Table~\ref{table:moti_dataset_comprison} summarizes the characteristics of a few popular publicly available software vulnerability datasets. We compare these datasets to highlight the contributions  \sysname can make.

Juliet~\cite{juliet}, Choi et.al~\cite{Choi_2017}, and S-babi~\cite{sbabi} are synthetic datasets that were generated from predefined patterns. Although their sizes are decent, the main drawback is the lack of diversity comparing to real-world programs \cite{Choi_2017}.

The examples in Draper \cite{draper} are from both synthetic and real-world programs, where each example contains a function and a few labels indicating the bug types. These labels were generated by aggregating static analysis results. Draper doesn't provide details like bug locations or traces. For real-world programs, Draper doesn't maintain the links to the original code base. If we want to further process the function-level examples to obtain more information, it's difficult to compile or analyze them without headers and compiler arguments.

The Devign \cite{devign} dataset contains real-world function examples from commits, where the labels are manually generated based on commit messages and code diffs. In particular, if a commit is believed to fix bugs, all functions patched by the commit are labeled as 1, which are not true in  many cases. In addition, only a small portion of the dataset was released.

CDG~\cite{vuldeepecker} is derived from real-world programs. It's unique because an example is a subset of a program slice and thus not a valid program. Its label was computed based on NVD: if the slice overlaps with a bug fix, it's labeled as $1$. Since the dataset is derived from confirmed bugs, the label quality is better. However, the number of such examples is limited and may not be sufficient for model training.

In fact, there is a pressing need for labeled datasets from real-world programs and encoding context information beyond the function boundary \cite{dl_vi_survey, devign, vuldeepecker}. It has been shown that preserving inter-procedural flow in code embedding can significantly improve the model performance (e.g. 20\% precision improvement in code classification task)~\cite{flow2vec}. To this end, \sysname examples are generated based on inter-procedural analysis, where an example can include multiple functions in the trace. \sysname also provides extra details such as the bug types, bug locations, bug traces, links to the original code base/commits, analyzer outputs, and compiler arguments that were used to compile the files having the functions. We believe they are helpful for AI for vulnerability detection in general.

\subsection{Manual Review and False Positive Reduction}
\label{sec:moti-opoenssl-manual-study}

We start by running a state-of-the-art static analyzer on a large real-world program. We select bug types that may lead to security problems and manually go through each issue to confirm how many reported issues are real bugs. 

The goal of this exercise is two-fold. First, we want to understand the performance of a state-of-the-art static analyzer for large real-world programs in terms of how many reported issues are real bugs. Second, by looking at the false positives, we want to explore ideas that can treat the static analyzer as a black box and suppress the false positives.

\subsubsection{Manual Case Study}

\begin{table}[t]
 	\centering
	\caption{Manual Review: OpenSSL \texttt{7f0a8dc}}
 	\vspace{-0.5em}
{\bgroup
\def\arraystretch{1.1}
\resizebox{0.9\columnwidth}{!}{%
    \begin{tabular}{l  r   r   r  r } 
      \toprule
        \multirow{2}{*}{\bf Error Type} &
        \multirow{2}{*}{\bf Reported} &
        \multicolumn{3}{c}{\bf Manual Review} 
      \\ \cmidrule(lr){3-5}
        & 
        &
        {\bf FP} & 
        {\bf TP} & 
        {\bf FP:TP} 
      \\ \midrule
        UNINITIALIZED\_VALUE & 
        101 & 
        101  &
        0 &
        --  
      \\ 
        NULL\_DEREFERENCE & 
        64 & 
        51  &
        13  &
        4:1
      \\ 
        RESOURCE\_LEAK & 
        1 & 
        1  &
        0  &
        --  
      \\ \hline
        {\bf TOTAL} & 
        {\bf 166} & 
        {\bf 153}  &
        {\bf 13}  &
        {\bf 12:1} \\ 
      \bottomrule
    \multicolumn{5}{l}{\textbf{\texttt{Note}}: 326 DEAD\_STORE issues were excluded from manual review. } \\
    \end{tabular}
    }
\egroup}
\label{table:opensll_manual}
\vspace{-0.1in}
\end{table}

\begin{figure}[t]
\begin{center}
\includegraphics[width=\columnwidth]{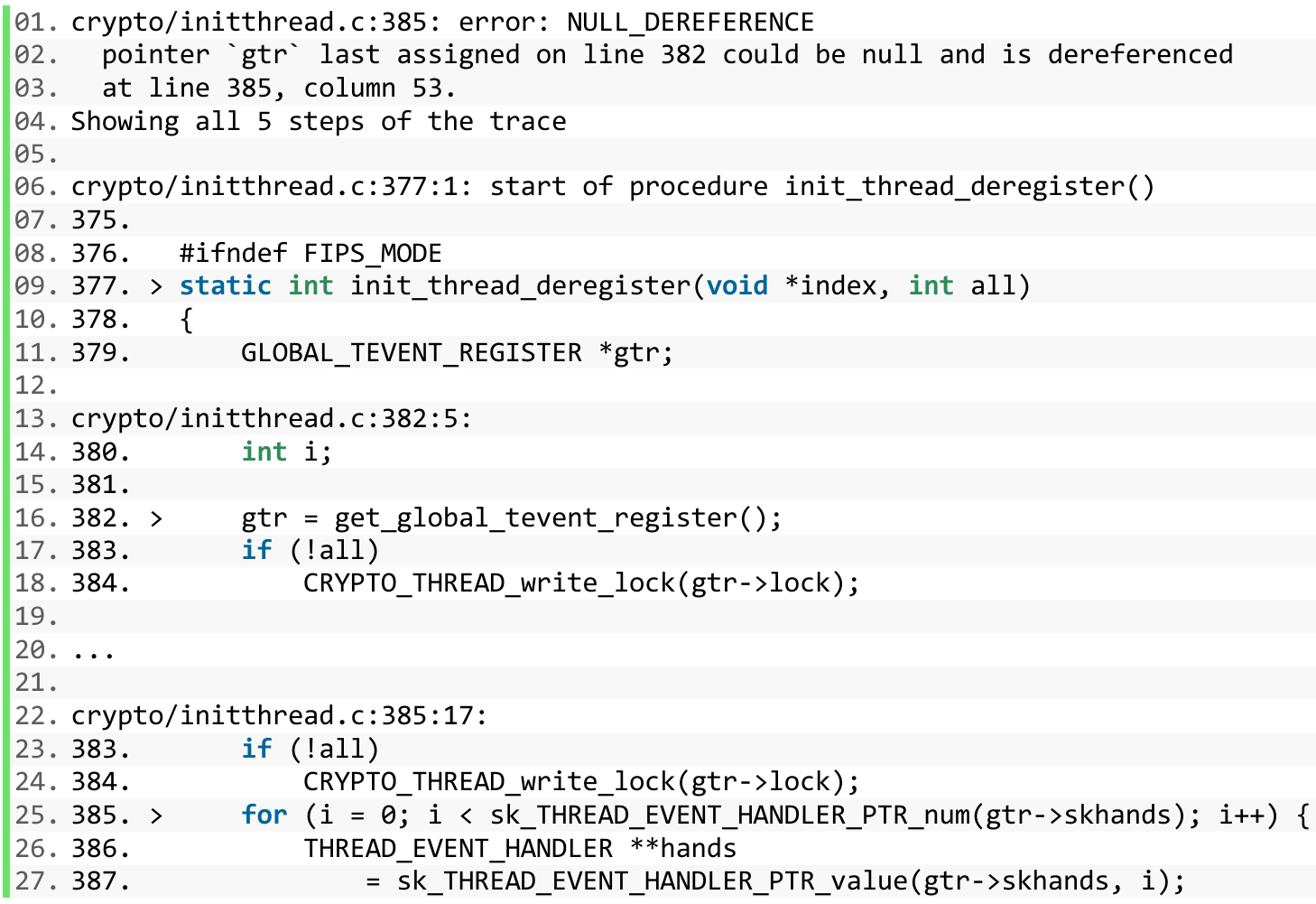}
\vspace{-0.25in}
\caption{Infer Bug Report Example.}
\label{fig:infer_report_eg}
\end{center}
\vspace{-0.25in}
\end{figure}

Since we are interested in large C/C++ programs, we require the static analyzer should be able to handle industrial-scale programs and detect a broad set of bug types. To the best of our knowledge, the Clang Static Analyzer \cite{csa} and Infer \cite{infer} are two state-of-the-art static analyzers that satisfy our needs. However, the Clang Static Analyzer doesn't support cross translation unit analysis such that the inter-procedural analysis may be incomplete. Therefore, we choose Infer in our experiments.  
We use \openssl version \texttt{7f0a8dc} as the benchmark, which has $1499$ \texttt{*.c}/\texttt{*.h} files and $513.6$k lines of C code in total. 

We run \infer using its default setting and the results are summarized in Table~\ref{table:opensll_manual}. \infer reported $492$ issues of $4$ bug types: $326$ \texttt{DEAD\_STORE}, $101$ \texttt{UNINITIALIZED\_VALUE}, $64$ \texttt{NULL\_DEREFERENCE}, and $1$ \texttt{RESOURCE\_LEAK}. Among them, \texttt{DEAD\_STORE} refers to the issues where the value written to a variable is never used. Since such issues are not directly related to security vulnerabilities, they were excluded from the manual review. The remaining $166$ issues may lead to security-related problems and thus were included in the study. 

The manual review was performed by 8 developers who are proficient in C/C++. We started by understanding the bug reports produced by \infer. Fig.~\ref{fig:infer_report_eg} shows an example of the bug report of a \texttt{NULL\_DEREFERENCE} issue. It has two sections. The bug location, bug type, and a brief justification why \infer thinks the bug can happen are listed in lines $1$--$3$. The bug explanation part can be in different formats for different bugs. In lines $6$-$27$, the bug trace that consists of the last steps of the offending execution is listed. Fig.~\ref{fig:infer_report_eg} shows 3 of the 5 steps. In each step (e.g. line $6$-$11$), the location and 4 additional lines of code that sit before and after the highlighted line are provided.  

We firstly had two rounds of manual analyses to figure out if the reported issue may be triggered. Each issue was reviewed by two reviewers. If both reviewers agreed that the reported bug can happen, we have an additional round of review and try to confirm the bug by constructing a test case. This process was very time consuming and challenging, especially when reviewing a complex program with cryptography involved. 

As shown in Table~\ref{table:opensll_manual}, out of 166 security vulnerability related issues, we confirmed that {\bf 13 (7.8\%)} issues are true positives and {\bf 92.2\%} are false positives. 



\subsubsection{Feature Exploration for False Positive Reduction}

\begin{figure}[t]
\begin{center}
\includegraphics[width=0.95\columnwidth]{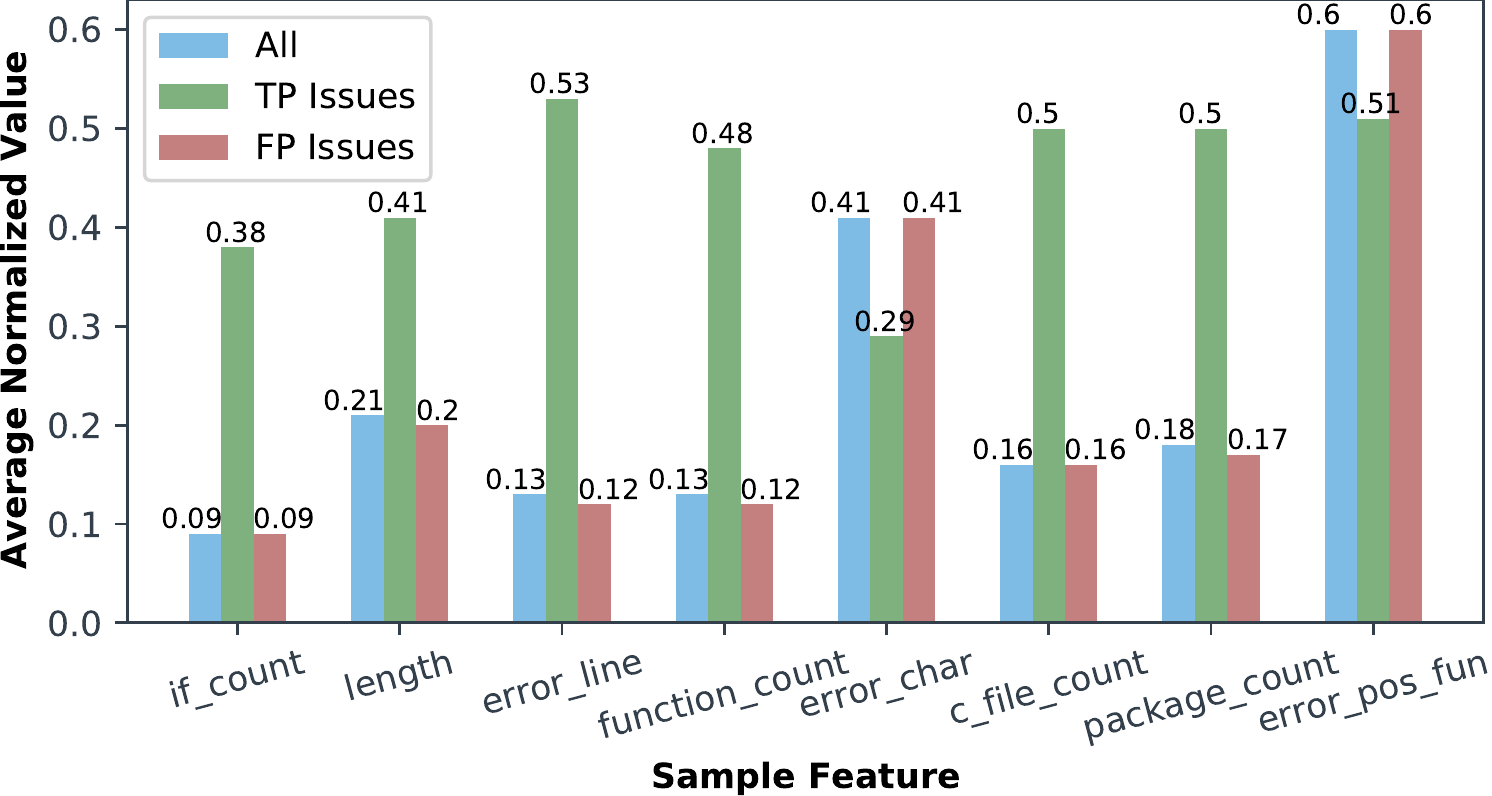}
\vspace{-0.1in}
\caption{Feature Exploration. We experiment with a few features that may reflect the complexity of the issues. After normalization, the averaged feature values of true positives and false positives are significantly different, which suggests a classifier may achieve good performance.}
\label{fig:openssl_manual_review_features}
\end{center}
\vspace{-0.25in}
\end{figure}

During the manual review, we found we can make a good guess for some issues by looking at the bug reports. Inspired by the existing false positive reduction works \cite{Yuksel_13, ALETHEIA}, we explored the idea of predicting if the issues flagged by \infer are true positives solely based on the bug reports as shown in Fig.~\ref{fig:infer_report_eg}. 

Existing approaches are not directly applicable as they target different languages or static analyzers. Following the intuition that complex issues are more likely to be false positives, we considered features in bug reports that may reflect the issue complexity.
We explored the following 8 features that belong to 3 categories: (1) {\it error\_line} and {\it error\_char} denote the location (line and column number) where the bug occurs. (2) {\it length}, {\it c\_file\_count} and {\it package\_count} denote the unique number of line numbers, source files and the directories respectively in the trace. (3) {\it if\_count} and {\it function\_count} are the numbers of the branches and functions in the trace.

We extracted the features from the bug reports of $166$ issues. After normalization, we computed the average feature values of the $13$ true positive issues and $153$ false positive issues. As shown in Fig.~\ref{fig:openssl_manual_review_features}, the average feature values of true positives and false positives are significantly different and easily separable for all 8 features, which suggests a good false positive reduction classifier can perform very well.

\section{Dataset Generation}
\label{sec:dataset}

In this section, we present the differential analysis based approach that labels the issues detected by the static analyzer. Then, we show how we generate two kinds of examples for the \sysname dataset based on the results obtained.  

\subsection{Overview}

\begin{figure*}[t]
\begin{center}
\includegraphics[width=0.9\textwidth]{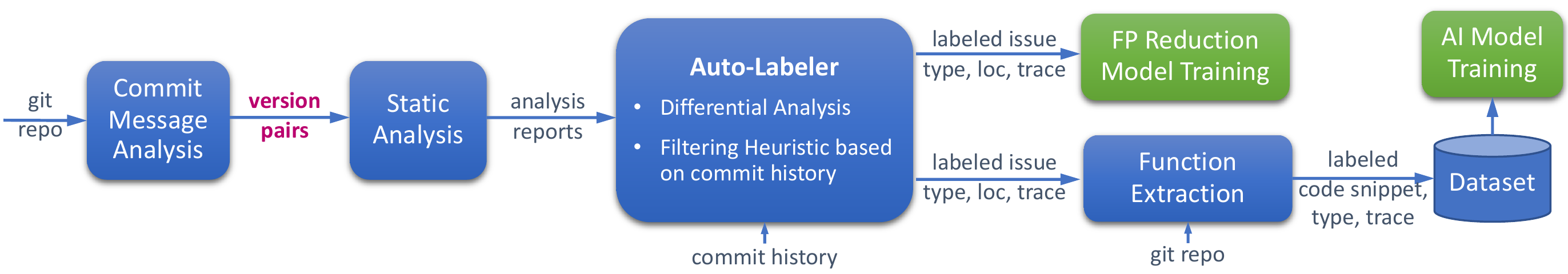}
\vspace{-0.1in}
\caption{The Overview of \sysname Dataset Generation Pipeline.}
\label{fig:pipeline_overviews}
\end{center}
\vspace{-0.2in}
\end{figure*}

Fig.~\ref{fig:pipeline_overviews} shows the overall workflow of \sysname. The input to the pipeline is a URL to a git repository. The output are examples generated purely using the static differential analysis. 

As the pre-processing step, based on the commit messages only, the Commit Message Analyzer (Sec.~\ref{sec:cma}) selects a list of commits that are likely to be bug fixes. Because it can be very expensive to analyze a pair of consecutive versions, the goal of this step is to filter out commits that are not closely related to bug fixes (e.g. documentation improvement commits) and speed up the process.  

For each selected commit, we obtain two sets of issues reported by the static analyzer by running the analyzer on the before-commit and the corresponding after-commit versions. The auto-labeler (Sec.~\ref{sec:auto-labeler-eg}) compares these two sets and identifies the issues that are fixed by the commit. 


After aggregating all such issues from multiple consecutive version pairs and filtering out noises based on commit history, the auto-labeler labels issues that are {\it very likely to be real bugs} as positives, and the issues that are never fixed by a commit as negatives because they are {\it very likely to be false positives}.  We further extract the function bodies according to the bug traces and create the dataset.



\subsection{Commit Message Analysis}
\label{sec:cma}

We created the Commit Message Analyzer (CMA) to identify commits that are more likely to refer to vulnerability fixes and not documentation changes or new features. Using the NVD dataset~\cite{NVD}, CMA learns the language of vulnerabilities and uses a hybrid approach that combines semantic similarity-based methods \cite{chandrasekaran2020evolution} and snippet samples-based methods \cite{snippet_similarity_06} to identify relevant commit messages and their associated commit. Noise is reduced by eliminating meaningless tokens, names, email addresses, links, code, etc from each commit message prior to the analysis.

Based on the semantic distribution of the vulnerable mentions, CMA identifies the category of the vulnerability and ranks commits based on confidence scores.


 

\subsection{Auto-labeler Examples}
\label{sec:auto-labeler-eg}

For each bug-fixing commit selected by CMA, we run the static analyzer on the versions before and after the commit. We evaluated several static analyzers such as CppCheck~\cite{cppcheck}, Flawfinder~\cite{flawfinder}, Clang Static Analyzer~\cite{csa}, and Infer~\cite{infer}. We chose Infer because it can detect a nice set of security related bug types and supports cross translation unit analysis necessary for effective inter-procedural analysis. More importantly, it scales well on large programs. 

\smallskip

\noindent
{\bf Identify Fixed Issues in a Version Pair.}
If we denote the issues found in the before-commit version as $I_{\text{before}}$  and the ones in the corresponding after-commit version as $I_{\text{after}}$, all issues can be classified into three groups: (1) the {\it fixed issues} ($I_{\text{before}} - I_{\text{after}}$) that are detected in the before-commit version but disappear in the after-commit version, (2) the {\it pre-existing issues} ($I_{\text{after}} \cap I_{\text{before}}$) that are detected in both versions, and (3) the {\it introduced issues} ($I_{\text{after}} - I_{\text{before}}$) that are not found in the before-commit versions but detected in the after-commit version. We are particularly interested in the {\it fixed issues} because they are very likely to be bugs fixed by the commit. We use the \texttt{infer-reportdiff} tool~\cite{infer_report_diff} to compute them.  

Note that it's possible that a {\it fixed issue} is not a real bug as the static analyzer may make mistakes, e.g. omit an issue from the after-commit even though the code had not changed. In our experience, an important reason is that Infer can exhibit non-deterministic behaviors \cite{infer_nondeterminism_1}. And the non-determinism occurs more frequently when enabling parallelization~\cite{infer_nondeterminism_2}. In order to minimize the impact, we have to run \infer in single-threaded mode. However, this setting brings in performance challenges and it takes several hours to analyze a version pair. For example, on an IBM POWER8 cluster, it takes 5.3 hrs and 12 hrs to analyze a version pair of \openssl and \ffmpeg, respectively, in single-thread mode. As we will need to analyze thousands of version pairs, it's impractical to do so on a PC or a small workstation. Therefore, we addressed several technical challenges and parallelized the analysis to process more than a thousand version pairs simultaneously in a cluster.

\smallskip

\noindent
{\bf Merge Issues Based on Commit History}. After identifying fixed issues in each version pair, we merge and deduplicate the issues from all version pairs. In particular, we compute the sha1sum of the bug report after removing location-related contents (e.g, the file names, line numbers, etc) and use it as the id for deduplication. The reason why we remove location-related contents is that the same piece of code may be bumped to a different location by a commit, changing only the line numbers in the report. Then, we apply the following two heuristics to filter out the ones that are unlikely to be bugs based on the commit history.
\begin{itemize}[leftmargin=*, nolistsep, noitemsep]
    \item {\it Fixed-then-unfixed issues}: The same issue may appear in multiple version pairs. We sort all its occurrences based on the author date of the commit. If a fixed issue appears again in a {\it later} version, it's probably a false positive due to the mistake of the static analyzer. We change the labels of such cases and mark them as {\it negatives}. 
    
    \item {\it Untouched issues}:  We check which parts of the code base are patched by the commit. If the commit code diff doesn't overlap with any step of the bug trace at all, it's unlikely the issue is fixed by the commit but more likely to be a false positive reported by the static analyzer. We mark such cases {\it negatives} as well.
\end{itemize}

After applying the above filters, the remaining issues in the {\it fixed issues} group are labeled as positives (issues that are more likely to be buggy) and all other issues are labeled as negatives (issues that are more likely to be non-buggy).
We call these {\it auto-labeler examples}. Because auto-labeler examples are generated based on issues reported by \infer, they all have the infer bug reports.

\subsection{After-fix Examples}
\label{sec:after-fix-eg}

Due to the nature of vulnerabilities, the auto-labeler produces many more negatives than positives such that the dataset of auto-labeler examples is quite imbalanced. Given that the positive auto-labeler examples are assumed to be bugs fixed in the after-commit versions, extracting the corresponding fixed versions is another kind of negative examples, which we call {\it after-fix examples}. There are two benefits: (1)  Since each negative example corresponds to a positive example, the dataset of auto-labeler positive examples and after-fix negative examples is balanced. (2) The after-fix negative examples are closely related to the positive ones so that they may help models focus on the delta parts that fixed the bugs.
Note that the {\it after-fix} examples do not have a static analysis bug report because the issue does not appear in the after-commit version.

\subsection{An Example in the \sysname Dataset}

\begin{figure}[t]
\begin{center}
\includegraphics[width=0.95\columnwidth]{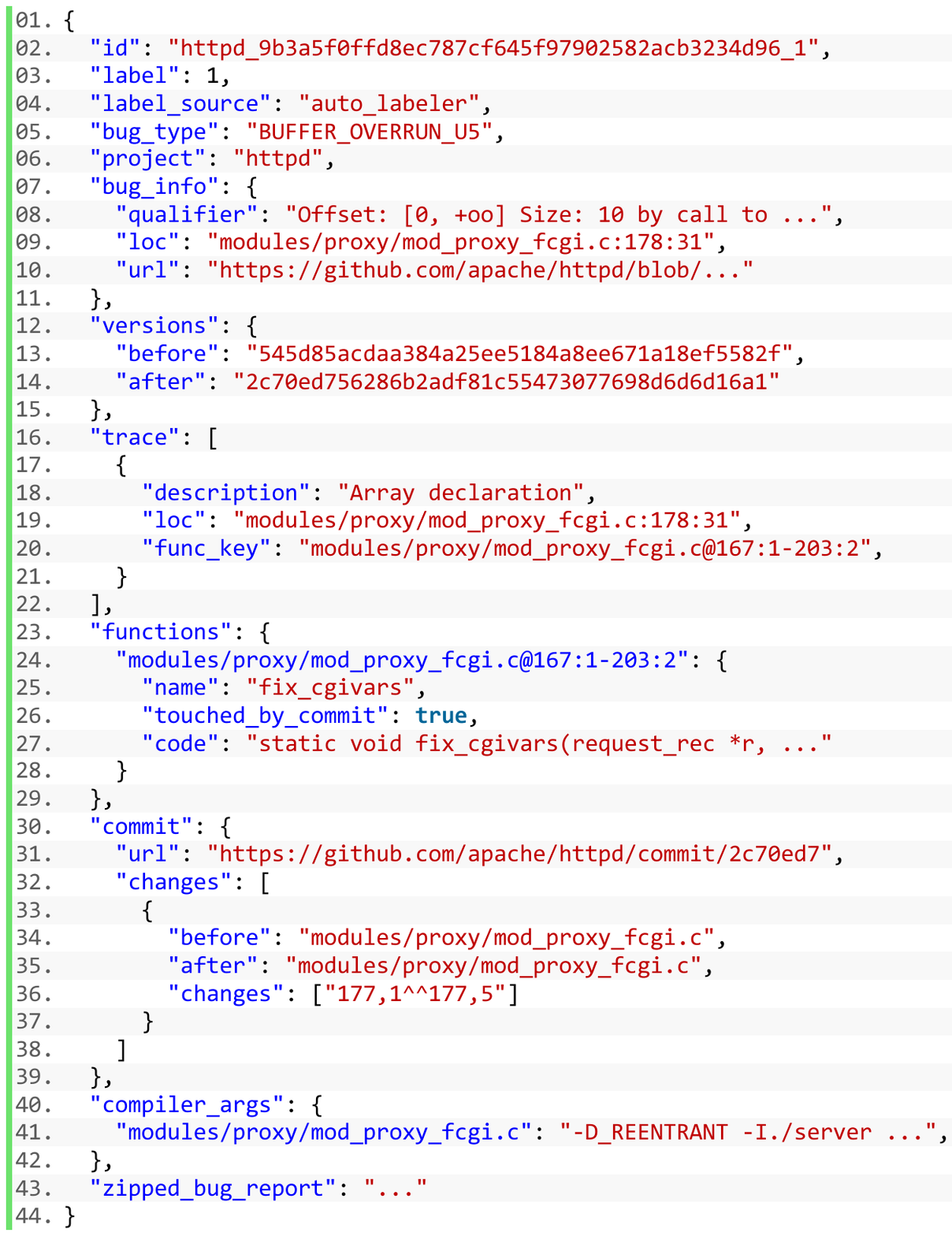}
\vspace{-0.1in}
\caption{A Simplified Example in \sysname Dataset.}
\label{fig:dataset_sample}
\end{center}
\vspace{-0.25in}
\end{figure}

Fig.~\ref{fig:dataset_sample} shows a \sysname example, which contains bug-related information obtained from the static analyzer, the code base, and the commit meta-data. 

In particular, every example has its {\it label} (0 or 1) and {\it label\_source} (``auto\_labeler" or ``after\_fix\_extractor") to denote how the example was generated and if it is buggy. {\it bug\_type},  {\it bug\_info}, {\it trace} and {\it zipped\_bug\_report} are obtained from the static analyzer, which provides details about the bug types, locations, traces, and the raw bug report produced by \infer. This information can be useful to train models on bug reports.  

For each step in the {\it trace}, if it refers to a location inside a function, we extract the function body and save it in the {\it functions} section. Therefore, an example has all functions involved in the bug trace, which can be used by function level or trace level models. Besides, we cross-check with commit code diff. If a function is patched by the commit, the {\it touched\_by\_commit} is true.

In addition, the compiler arguments used to compile the source file are saved in the {\it compiler\_args} field. They can be useful when we want to run extra analysis that requires compilation (e.g. libclang \cite{clang_tools} based tools). 

\section{Static Analysis False Positive Reduction}
\label{sec:design}

Although the \sysname dataset is mainly for machine learning based vulnerability detection methods, in this paper, we show the dataset can be used to train a static analysis false positive reduction model and help developers prioritize potential true positives. We will present AI-based code understanding approaches for vulnerability detection tasks in another paper. 

\subsection{Problem Statement}
\label{sec:fp_reduction_ps}

As observed previously \cite{Johnson_icse13, Muske_13}, an excessive number of false positives greatly hinders the utilization of static analyzers as developers get frustrated and do not trust the tools. To this end, we aim to device a method that can identify a subset of the reported issues that are more likely to be true positives, and {\it use it as a prioritization tool}. Developers may focus on the issues selected by the model first and then move to remaining issues with a higher false positive rate if they have time. 

We treat the static analyzer as a black box and train a false positive reduction model solely based on the bug reports. Our goal is to achieve a balance between a large number of predicted positives and a high false positive reduction rate. We want developers to see more real bugs in the predicted positives comparing to all issues reported by the static analyzer.

\subsection{Static Analysis Outputs/Data}
\paragraph{Bug Trace description}
Infer static analysis produces many output files. For our purposes, we are only interested in the bug trace text file, illustrated in Fig.\ref{fig:infer_report_eg}, from which we extract the features.
The bug trace starts with the location where the static analyzer believes the error to have originated, and lists all the steps up to the line generating the error.
Many of the bugs are inter-procedural, so the bug trace cuts across many files and functions. 
For each step in the flow, the trace contains 5 lines of code centered on the statement involved, the location of the file and function in the project, and a brief description of the step.
At the top of the trace, the file and line of code where the bug occurred are mentioned along with the bug type (error type). There is also a short description of the bug.
The bug trace is therefore a combination of different types of data like source code, natural language, numeric data like line numbers, and file paths.


\paragraph{Dataset description}
As described in Sec.\ref{sec:after-fix-eg}, the original dataset has two types of negative examples, before-fix and after-fix. For these experiments, we built a dataset using the positive samples and the before-fix negative examples.
We are not interested in the after-fix negative examples since these samples don't produce a bug trace.
In every project, the number of negative labels is very large compared to the number of positive labels, as can be seen in Table \ref{table:secuirty_realted_error_types}.


\subsection{Feature Engineering}

Our primary assumption when coming up with features was that complex code is more likely to have bugs and/or is more likely to be classified as having bugs by a static analyzer, because it is highly probable that the developer failed to  consider all possible implications of the code. 
Complex code is also more difficult for other developers to understand, increasing the chance of their introducing bugs.

One indication of complexity is the size of the bug trace. A long bug trace indicates that the control passes through many functions, files or packages. 
The location of the bug could also indicate the complexity of the code. The line number is indicative of the size of the file, and the column number indicates the length of the line of the code where the bug occurred. The depth of the line of code could indicate how entrenched the problematic code happens to be.
Conditional statements cause many branches of execution to emerge and these can lead to convoluted and buggy code. One way to estimate the complexity is to count the number of times conditional statements occur and also the occurrences of OR/AND conditions.
The error type is also a major feature that we consider, as well as the number of C keywords used.
Table~\ref{table:features} lists our final set of features. We extract and normalize these features and save them in a features file.


\begin{table*}[t]
\centering
	\caption{Features Extracted from Infer Bug Report}
 	\vspace{-0.5em}
{\bgroup
\def\arraystretch{0.95}
\resizebox{0.9\textwidth}{!}{%
    \begin{tabular}{p{2.9cm}  p{5cm}  p{2.9cm}  p{5cm}}
      \toprule
          \cellgrey {\bf Feature}  & {\bf Description} & 
          \cellgrey {\bf Feature}  & {\bf Description} \\
      \midrule
          \cellgrey error & Infer bug/issue type & 
          \cellgrey error\_line & line number of the error 
          \\
          \cellgrey error\_line\_len & length of error line & 
          \cellgrey error\_line\_depth & indent for the error line text 
          \\
          \cellgrey average\_error\_line\_depth & average indent of code lines &
          \cellgrey max\_error\_line\_depth & max indent of code lines
          \\ 
          \cellgrey error\_pos\_fun & position of error within function & 
          \cellgrey average\_code\_line\_length & average length of lines in flow 
          \\
          \cellgrey max\_code\_line\_length & max length of lines in flow & 
          \cellgrey length & the number of lines of code 
          \\
          \cellgrey code\_line\_count & the number of flow lines & 
          \cellgrey alias\_count & the number of address assignment lines 
          \\
          \cellgrey arithmetic\_count & average operators / step & 
          \cellgrey assignment\_count & fraction of Assignment steps 
          \\
          \cellgrey call\_count & fraction of {\tt call} steps & 
          \cellgrey cfile\_count & the number of different .c files 
          \\
          \cellgrey for\_count & the number of {\tt for} loops in report & 
          \cellgrey infinity\_count & fraction of +00 steps 
          \\
          \cellgrey keywords\_count & the number of C keywords & 
          \cellgrey package\_count & the number of different directories 
          \\
          \cellgrey question\_count & fraction of '??' steps & 
          \cellgrey return\_count & average branches / step 
          \\
          \cellgrey size\_calculating\_count & average size calculations / step & 
          \cellgrey parameter\_count & fraction of parameter steps 
          \\
          \cellgrey offset\_added & the number of ``offset added"s in report & 
          \cellgrey  &  \\
      \bottomrule
    \end{tabular}
    }
\egroup}
\label{table:features}
\vspace{-0.1in}
\end{table*}

\subsection{Model Selection}

We experimented with 13 well-known machine learning models: namely, Decision Trees, K-means, Random Forest, Extra-trees, Gradient Boosting, Ada Boost, XGBoost, Catboost, LightGBM, Linear Classifiers with Stochastic Gradient Descent, Gaussian Naive Bayes, Multinomial Naive Bayes, and Complement Naive Bayes.
We ranked them based on both their AUC and F1 scores and selected the four best models.  These were based on an ensemble of decision trees, Random Forest and Extra Trees for bagging methods, LightGBM, and Catboost for boosting.

Random Forest is made of many weak learners (single decision trees), which are fed with random samples of the data and trained independently using a random sample of the features.
Each inner decision tree is grown by using the features which offer the best split at every step. The randomness, which makes every single tree different from the rest of the forest, associated with a high number of learners, makes the model quite robust against overfitting. 
A slightly different variation of Random Forest is Extra-Trees, also known as Extremely Randomized Trees. The only difference is how the data is sampled to create the input and how the splits are chosen randomly making the forest more diversified. 
Differently from bagging methods, where learners are trained independently, with gradient boosting methods each tree improves over the predictions made by the previous ones. Boosting techniques are known to work successfully with imbalanced data, but might suffer more from overfitting - to mitigate this effect  typically many trees are used together. LightGBM and Catboost are different frameworks to implement this kind of ensemble: the former creates an imbalanced tree while the latter creates a balanced one.

\subsection{Evaluation Metrics}

\begin{figure}
\centering
\includegraphics[width=0.85\columnwidth]{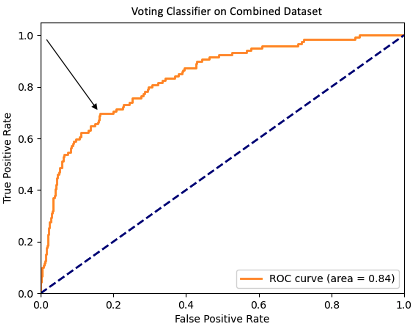}
\vspace{-0.15in}
\caption{The ROC curve tracks the performance at all classification thresholds, while the Area under the Curve (AUC) provides an aggregation of performance across all possible classification thresholds. The point on the curve that minimizes the distance from the top-left corner is the one which provides the best compromise between false positive rate and true positive rate.}
\vspace{-0.2in}
\label{fig:randomForest_ROC_AUC}
\end{figure}

In order to evaluate the different models, because of the imbalance in the dataset, we used the Area Under the Curve (AUC score, Fig.~\ref{fig:randomForest_ROC_AUC}), a threshold-invariant metric that visualizes the trade-off when we want to reduce the false positive rate while maintaining a good true positive rate. 
Since the main task is to reduce the number of False Positives, we calculate the percentage reduction in False Positives on the test set. Relying too much on this metric can bias towards models, which make very few accurate predictions. 
To make sure this is not the case, we also calculate the total percentage of True Positives which are predicted by the model. 
An ideal model would have a very high AUC Score, low False Positive rate, and high True Positives rate: to choose the threshold for the AUC, we select that point which minimizes the distance from the top-left corner (all true positives and no false positives). Once this point is chosen we also present F1-score as the average of each class F1-score since our goal is to reduce the number of false positives while preserving the real ones. 

\subsection{Voting}
Real-world datasets present a high imbalance between real bugs and false positives. Also, the projects used to derived the datasets proposed in this work vary in size yielding different dataset sizes. Therefore, it's not easy to choose the model which does the best on all the datasets. While on a specific dataset a model can perform greatly, it could work poorly in another: to mitigate such a problem we applied a soft-voting strategy, which combines the scores of each classifier, which should guarantee a more stable behavior across datasets.

\section{Evaluation}
\label{sec:eval}

In this section, we present the result of \sysname dataset generation and label evaluation. In addition, we show the evaluation results of the AI-based static analysis false positive reduction as a use case to demonstrate how \sysname dataset can be helpful.

\subsection{Dataset Generation Results}

\subsubsection{Dataset Statistics}

\begin{table*}[th]
\centering
	\caption{Dataset Generation Results}
 	\vspace{-0.5em}
{\bgroup
\setlength{\tabcolsep}{8pt}
\def\arraystretch{0.9}
\resizebox{0.8\textwidth}{!}{%
    \begin{tabular}{l r r  r r r r r} 
      \toprule
        \multirow{2}{*}{\bf Project} &
        \multicolumn{2}{c}{\bf Version Pairs} &
        \multirow{2}{*}{ \makecell{ {\bf Issues} \\ {\bf Reported}} }  &
        \multicolumn{3}{c}{{\bf Unique Auto-labeler Examples}} & 
        \multicolumn{1}{c}{{\bf Unique After-fix Examples}} 
      \\ \cmidrule(lr){2-3}
         \cmidrule(lr){5-7}
         \cmidrule(lr){8-8}
       &
       {\bf CMA} &
       {\bf Infer} &
       &
        \multicolumn{1}{c}{\bf All} &
        \multicolumn{1}{c}{\bf Negatives} &
        \multicolumn{1}{c}{\bf Positives} & 
     \multicolumn{1}{c}{\bf Negatives}
      \\ \midrule 
        \openssl & 3,011 & 2,643 & 42,151,595  & \cellgreen 351,170 & \cellgreen 343,148 & \cellgreen 8,022   & \cellblue 8,022
      \\
        \ffmpeg  & 5,932 & 4,930 & 215,662,372 & \cellgreen 659,717 & \cellgreen 654,891 & \cellgreen 4,826   & \cellblue 4,826
      \\
        \httpd   & 1,168 & 542   &   1,681,692 & \cellgreen 12,692 & \cellgreen 12,475  & \cellgreen 217     & \cellblue 217
      \\
        \nginx   & 785   & 635   & 3,283,202   & \cellgreen 18,366 & \cellgreen 17,945  & \cellgreen 421      & \cellblue 421
      \\
        \libtiff & 144   & 144   & 525,360     & \cellgreen 12,649 & \cellgreen 12,096  & \cellgreen 553       & \cellblue 553
      \\ 
        \libav   & 3,407 & 2,952 & 86,069,532  & \cellgreen 241,029 & \cellgreen 236,415 & \cellgreen 4,614    & \cellblue 4,614
      \\ \midrule
        {\bf Total} & {\bf 14,447} & {\bf 11,846} & {\bf 349,373,753} & \cellgreen {\bf 1,295,623} & \cellgreen {\bf 1,276,970} & \cellgreen {\bf 18,653} & \cellblue {\bf 18,653}  \\
      \bottomrule

      \multicolumn{8}{l}{ 
      \begin{minipage}{0.7\textwidth}
            \smallskip
            \begin{itemize}[leftmargin=*]
                \item {\bf CMA:} The number of bug-fixing commits identified by the commit message analyzer.
                \item  {\bf Infer:} The number of version pairs successfully analyzed by Infer.
                \item  {\bf Issues Reported:} The number of issues detected in the before-commit versions before deduplication. 
            \end{itemize}
        \end{minipage}
        }
    \end{tabular}
    }
\egroup}
\label{table:dataset_gen_stats}
\vspace{-0.15in}
\end{table*}

The dataset generation pipeline is written in python and runs on a POWER8 cluster, where each node has 160 CPU cores and 512GB RAM. We analyzed 6 open-source programs (namely, \openssl, \ffmpeg, \httpd, \nginx, \libtiff, and \libav) and generated the initial version of the \sysname dataset. 
In particular, \infer can detect more than 150 types of issues in C/C++/Objective-C/Java programs \cite{infer_issue_types}. However, some issues are not ready for production and thus disabled by default. In the pipeline, we additionally enabled all issue types related to buffer overflows, integer overflows, and memory/resource leaks, even some of them may not be production-ready. 

Table~\ref{table:dataset_gen_stats} summarizes the dataset generation results. The column {\it CMA Version Pairs} shows the number of bug-fixing commits selected by the commit message analyzer (Sec. \ref{sec:cma}). For each selected commit, we run \infer on both the before-commit and after-commit versions. We drop a commit if \infer failed to analyze either the before-commit version or the after-commit version. Column {\it \infer} shows the number of commits or version pairs \infer successfully analyzed. For auto-labeler examples (Sec. \ref{sec:auto-labeler-eg}), column {\it Issues Reported} and {\it unique auto-labeler examples - all} shows the number issues \infer detected in the before-commit versions before and after deduplication, which will be labeled as positives and negatives as shown in column {\it Positives} and {\it Negatives}. For after-fix examples (Sec. \ref{sec:after-fix-eg}), column {\it Negatives} shows the number of examples generated based on the auto-labeler positive examples. In total, we processed 11,846 consecutive versions pairs. Based on the results, we generated 1,295,623 unique auto-labeler examples and 18,653 unique after-fix examples.

\subsubsection{Manual Label Validation}

\begin{table}[t]
\centering
	\caption{Auto-labeler Manual Validation Results}
 	\vspace{-0.5em}
\addtolength{\tabcolsep}{-4pt} 
{\bgroup
\def\arraystretch{0.9}
\resizebox{\columnwidth}{!}{%
    \begin{tabular}{l c c c c c c c c c} 
      \toprule
        \multirow{2}{*}{} &
        \multicolumn{3}{c}{\bf Positives} &
        \multicolumn{3}{c}{\bf Negatives} & 
        \multicolumn{3}{c}{\bf All} 
      \\ \cmidrule(lr){2-4}
         \cmidrule(lr){5-7}
         \cmidrule(lr){8-10}
       &
       {\bf \#} &
       \cellgreen {\bf A} &
       {\bf D} &
       {\bf \#} &
        \cellgreen {\bf A} &
       {\bf D} &
       {\bf \#} &
       \cellgreen {\bf A} &
       {\bf D}
      \\ \midrule 
        BUFFER\_OVERRUN\_L1     &  2  &\cellgreen  0  &  2    
                                &  1  &\cellgreen  1  &  0 
                                &  3  &\cellgreen  1  &  2  \\
                                
        BUFFER\_OVERRUN\_L2     &  3  &\cellgreen  1  &  2    
                                &  1  &\cellgreen  1  &  0 
                                &  4  &\cellgreen  2  &  2  \\
                                
        BUFFER\_OVERRUN\_L3     &  6  &\cellgreen  1  &  5    
                                &  4  &\cellgreen  4  &  0 
                                & 10  &\cellgreen  5  &  5  \\
                                
        BUFFER\_OVERRUN\_S2     &  0  &\cellgreen  0  &  0    
                                &  1  &\cellgreen  0  &  1 
                                &  1  &\cellgreen  0  &  1  \\
                                
        INTEGER\_OVERFLOW\_L1   &  3  &\cellgreen  2  &  1    
                                &  1  &\cellgreen  1  &  0 
                                &  4  &\cellgreen  3  &  1  \\   
                                
        INTEGER\_OVERFLOW\_L2   & 13  &\cellgreen  6  &  7    
                                &  3  &\cellgreen  3  &  0 
                                & 16  &\cellgreen  9  &  7  \\  
                                
        INTEGER\_OVERFLOW\_R2   &  1  &\cellgreen  1  &  0    
                                &  0  &\cellgreen  0  &  0 
                                &  0  &\cellgreen  1  &  0  \\ 
                                
        MEMORY\_LEAK            &  1  &\cellgreen  1  &  0    
                                &  1  &\cellgreen  1  &  0 
                                &  2  &\cellgreen  2  &  0  \\

        NULL\_DEREFERENCE       &  2  &\cellgreen  1  &  1    
                                &  1  &\cellgreen  0  &  1 
                                &  3  &\cellgreen  1  &  2  \\
                                
        RESOURCE\_LEAK          &  1  &\cellgreen  1  &  0    
                                &  1  &\cellgreen  1  &  0 
                                &  2  &\cellgreen  2  &  0  \\

        UNINITIALIZED\_VALUE    &  9  &\cellgreen  3  &  6    
                                &  1  &\cellgreen  1  &  0 
                                & 10  &\cellgreen  4  &  6  \\

        USE\_AFTER\_FREE        &  0  &\cellgreen  0  &  0    
                                &  1  &\cellgreen  1  &  0 
                                &  1  &\cellgreen  1  &  0  \\

      \midrule
        ALL                     & \makecell{41 \\ 100\% }  &\cellgreen \makecell{17 \\ 41\% }  &  \makecell{24 \\ 59\%}    
                                & \makecell{16 \\ 100\% }  &\cellgreen \makecell{13 \\ 81\% }  &  \makecell{ 3 \\ 19\%}  
                                & \makecell{57 \\ 100\% }  &\cellgreen \makecell{30 \\ 53\% }  &  \makecell{27 \\ 47\%} \\
                                
      \bottomrule
      \multicolumn{10}{l}{ 
      \begin{minipage}{1.1\columnwidth}
            \smallskip
            \begin{itemize}[leftmargin=*]
                \item {\bf \#:} the issue count; {\bf A/D:} manual review agrees/disagrees with the auto-labeler label
            \end{itemize}
        \end{minipage}
      }
    \end{tabular}
    }
\egroup}
\label{table:label_manual_validation}
\vspace{-0.1in}
\end{table}

As there is no ground truth, to evaluate the label quality we randomly selected 57 examples (41 positives, 16 negatives) with a focus on positives. We gave more weights to positive examples because they are more important for our purpose. As mentioned in Sec.~\ref{sec:fp_reduction_ps}, labeling a non-buggy example as buggy is against the goal of false positive reduction. But it's acceptable if we miss some of the real bugs. If we select examples according to the overall dataset distribution, we will have too few positive examples. 
Each example was independently reviewed by 2 reviewers.

Table~\ref{table:label_manual_validation} shows the label validation results. On this biased sample set, the accuracy with and without the auto-labeler is 53\% and 35\% respectively. Note the accuracy on an unbiased sample set is expected to be higher as there should be more negative examples. Take the \openssl study in Sec.~\ref{sec:moti-opoenssl-manual-study} as an example. Without auto-labeler, the accuracy was only 7.8\% on the set of 166 security-related examples.

\subsection{False Positive Reduction Results}


\subsubsection{Dataset}

To facilitate reproducibility, we defined and plan to release a split for each project. In particular, we drop bug types without any positive examples and split each project's data into {\it train:dev:test} sets (80:10:10) while maintaining the distribution of bug types. We use the same split in this experiment. The model will be trained on the {\it train} + {\it dev} sets and tested on the {\it test} set. 

We observed that some \ffmpeg and \libav examples are quite similar as \libav was forked from \ffmpeg \cite{libav_ffmpeg}. We dropped \ffmpeg examples so that the all-data combined experiment would be fair. \ffmpeg examples are more imbalanced compared to \libav and we leave it for future work.

Although we collect examples generated for many bug types that are not production-ready and are disabled by default in \infer, in this experiment, we consider just the 18 {\it security-related bug types} that are enabled by default. Table~\ref{table:secuirty_realted_error_types} shows the statistics of the data used in the experiment.

\begin{table}[t]
 	\centering
	\caption{Production-ready Security Related Error Types Filtering}
 	\vspace{-0.5em}
{\bgroup
\def\arraystretch{0.9}
\resizebox{1\columnwidth}{!}{%
    \begin{tabular}{l  r  r  r   r  r  r} 
      \toprule
        &
        \multicolumn{3}{c}{\bf All Errors} & 
        \multicolumn{3}{c}{\bf Prod-ready Sec Errs} 
      \\ 
      \cmidrule(lr){2-4}
      \cmidrule(lr){5-7}
      &
      {\bf Negatives} & {\bf Positives} & {\bf N:P} &
      {\bf Negatives} & {\bf Positives} & {\bf N:P} \\
      \midrule
      \openssl &  341,625  &   7,916  &  43:1 &  27,227  &  797  &  34:1 \\
      \libav   &  235,369  &   4,585  &  51:1 &  14,954  &  280  &  53:1 \\
      \nginx   &   1,7829  &     417  &  43:1 &   1,446  &   36  &  40:1 \\
      \libtiff &   11,720  &     552  &  21:1 &   1,185  &   27  &  44:1 \\
      \httpd   &   11,511  &     208  &  55:1 &     174  &   11  &  16:1 \\  
       \bottomrule
    \end{tabular}
    }
\egroup}
\label{table:secuirty_realted_error_types}
\vspace{-0.1in}
\end{table}

\subsubsection{Feature Importance}
\begin{figure}[t]
\begin{center}
\includegraphics[width=0.9\columnwidth]{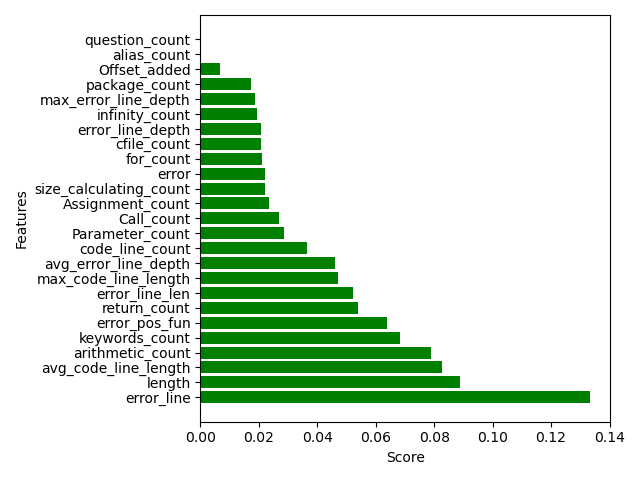}
\vspace{-0.2in}
\caption{Feature Importance of Random Forest algorithm trained on OpenSSL}
\label{fig:OpenSSLAvgFeat}
\end{center}
\vspace{-0.25in}
\end{figure}

We used feature importance ranking when selecting the final set of 25 features.  Figure~\ref{fig:OpenSSLAvgFeat} shows the features and their relative importance for one of the models. Two features that were important for many models are the line number of the error in the file and the number of lines of code in the bug report, perhaps suggesting that large files and complex bug reports distinguish real errors. 

\subsubsection{FP Reduction Model}

We trained Random Forest using 1000 estimators, Extra Trees with 500. For the Boosting algorithms, we used 500 estimators, learning rate 0.03, importance type \textit{gain} for the LGBM classifier, and the same number of estimators for Catboost. 
We define False Positive Reduction Rate (FPRR) = $(\text{FP}_{\text{infer}}$ - $\text{FP}_{\text{predict}}) / \text{FP}_{\text{infer}} * 100$. Because all issues are positive according to \infer, $\text{FP}_{\text{infer}}$ is just the number of negative examples. 

As shown in Table~\ref{table:self_pred_result}, 
all models can effectively reduce the false positives for each project. In most cases, the FPRR is above 70\% for every model, without penalizing too much the reduction of true bugs. As expected, it's hard to find one best model across all the projects. However, it's encouraging that the voting can outperform the single models in the \combined experiment, letting us believe that the more data we have, the better the voting system could perform.

\begin{table}[t]
 	\addtolength{\tabcolsep}{-3pt} 
	\caption{False Positive Reduction Results} 
 	\vspace{-0.5em}
{\bgroup
\def\arraystretch{0.9}
\resizebox{1\columnwidth}{!}{%
    \begin{tabular}{l l  r r r   r r r   r r  r   r r r} 
      \toprule
        &
        {\bf Model} &
        {\bf GP} & {\bf P} & {\bf TP} & 
        {\bf GN} & {\bf N} & {\bf TN} & 
        {\bf F1} &\cellgreen {\bf FPRR} & {\bf AUC}
      \\ \midrule
        \multirow{5}{*}{\openssl} &
        cb &
            81   & 858  & 62 &
            2711   & 1934 & 1915  & 
            0.48   &\cellgreen 70.6\%    & 0.79 
      \\
        &
        lgbm &
            81   & 827 & 65 &
            2711   & 1965 & 1949  & 
            0.49   &\cellgreen  71.9\%    & 0.82
        \\
        &
        rf &
            81   & 591  & 59 &
            2711   & 2201 & 2179  & 
            0.53   &\cellgreen  80.4\% & 0.83
        \\
        &
        etc &
            81 & 616  &52 &  
            2711  & 2176 & 2147 & 
            0.51  &\cellgreen  79.2\%	& 0.78
        \\
        \cmidrule{2-11}
        &
         voting & 
            81 & 506 & 58 &
            2711 & 2286 & 2263  & 
            0.55 &\cellgreen  83.5\%    & 0.83
        \\ 
      \midrule
      
        \multirow{5}{*}{\libav} &
        cb &
            28   & 256  & 22  &
            1495   & 1266  & 1260  & 
            0.53   &\cellgreen  84.3\%    & 0.89 
      \\
        &
        lgbm &
            28   & 220  & 21 &
            1495   & 1303 & 1296  & 
            0.55   &\cellgreen  86.7\%    & 0.91 
        \\
        &
        rf &
            28   & 287  & 22 &
            1495   & 1236 & 1230  & 
            0.52   &\cellgreen  82.3\%    & 0.87  
        \\
        &
        etc &
            28 & 54  &  13 &
            1495  & 1469  & 1454 & 
            0.65  &\cellgreen  97.3\%	& 0.70
        \\
        \cmidrule{2-11}
        &
         voting &
            28 &  254 & 21 & 
            1495 & 1269 & 1262 &  
            0.53 &\cellgreen 84.4\% & 0.89
          \\ 
     \midrule

        \multirow{5}{*}{\nginx} &
        cb &
            5   &  27 & 3 &
            145   & 123 & 121  & 
            0.55   &\cellgreen  83.4\%    &   0.85

        \\
        &
        lgbm & 
            5   & 47  & 4 &
            145   & 103 & 102  & 
            0.49   &\cellgreen  70.3\%    & 0.86
        \\
        &
        rf &
            5   & 46  & 3 &
            145   & 104 & 102  & 
            0.47   &\cellgreen  70.3\%    & 0.75 
        \\
        &
        etc &
            5 & 60 & 3 &
            145  & 90 & 88 & 
            0.42  &\cellgreen  60.7\%	& 0.67 
        \\
        \cmidrule{2-11}
        &
         voting &

            5 & 54  & 4 &
            145 & 96 & 95 & 
            0.46  &\cellgreen  65.5\%    &  0.78
          \\ 
      \midrule
      
        \multirow{5}{*}{\libtiff} &
        cb &
            3   & 17  & 2  &
            118  & 104 & 103  & 
            0.56   &\cellgreen  87.3\%    & 0.92
      \\
        &
        lgbm &
            3  & 5 & 1 &
            118 & 116 & 114 & 
            0.61   &\cellgreen  96.6\%    & 0.72  
        \\
        &
        rf &
            3   & 7  & 2 &
            118   & 114 & 113 & 
            0.69   &\cellgreen  95.8\%    &  0.98 
        \\
        &
        etc &
            3 & 8 & 2 &
            118  & 113 & 112 & 
            0.67  &\cellgreen  94.9\%	& 0.97
        \\
        \cmidrule{2-11}
        &
         voting &
            3  & 7 & 2  &
            118   & 114 &  113 & 
            0.69   &\cellgreen    95.8\%  & 0.97
          \\ 
      \midrule
      
        \multirow{5}{*}{\httpd} &
        cb &
            2 & 5 & 1 &
            17  & 14  & 13 & 
            0.56 &\cellgreen 76.5\%	& 0.88
      \\
        &
        lgbm &
            2 & 3 & 1 &
            17  & 16 & 15 & 
            0.65 &\cellgreen 88.2\%	& 0.94
        \\
        &
        rf &
            2 & 9 & 1 &
            17  & 10 & 9 & 
            0.42 &\cellgreen 52.9\%	& 0.77
        \\
        &
        etc &
            2 & 6 & 1 &
            17  & 13 & 12 & 
            0.53 &\cellgreen 70.6\%	& 0.85
        \\
        \cmidrule{2-11}
        &
         voting &
            2  & 6  & 1 &
            17 & 13 & 12  & 
            0.53 &\cellgreen 70.6\% &  0.85
          \\ 
      \midrule

        \multirow{5}{*}{\combined} &
        cb &
            119   & 1403  & 95  &
            4486   & 3202  & 3178  & 
            0.48   &\cellgreen  70.8\%    & 0.82 
      \\
        &
        lgbm &
            119   & 1274  & 93 &
            4486   & 3331 & 3305  & 
            0.49  &\cellgreen  73.7\%    & 0.83  
        \\
        &
        rf &

            119   & 1063 & 86 &
            4486   & 3542 & 3509  & 
            0.50  &\cellgreen  78.2\%    & 0.84
        \\
        &
        etc &
            119 & 1053  & 74 &
            4486  & 3552  & 3507 & 
            0.50  &\cellgreen  78.2\%	& 0.74 
        \\
        \cmidrule{2-11}
        &
         voting &
            119 &  814 & 82 & 
            4486 & 3791 & 3754 &  
            0.54 &\cellgreen 83.7\% & 0.84 
          \\ 
      \bottomrule
    \multicolumn{11}{l}{
        \begin{minipage}{1.05\columnwidth}
            \smallskip
            \smallskip
            \begin{itemize}[leftmargin=*]
                \item The released dataset split defines train/dev/test for each project. For combined, its train/dev/test sets are the union of corresponding sets of all project
                \item The models are trained on train + dev sets and tested on the test set.
                \item  {\bf GP/P/TP} Ground-truth/Predicted/True Positives; {\bf GN}/{\bf N}/{\bf TN} defined similarly.
                \item {\bf FPRR}: False Positive Reduction Rate = {\bf (GN - FP) / GN * 100}   
                \item {\bf cb}: Catboost, {\bf lgbm}: LightGBM, {\bf rf}: Random Forest, {\bf etc}: Extra-Trees
            \end{itemize}
        \end{minipage}
    }
    

    
    \end{tabular}

    }
\egroup}
\label{table:self_pred_result}
\vspace{-0.10in}
\end{table}

\section{Related Work}
\label{sec:related}

\noindent
\textbf{Datasets for AI-based Vulnerability Detection}. Juliet~\cite{juliet},  Choi  et.al~\cite{Choi_2017}, and S-babi~\cite{sbabi}  are synthetic datasets. They are generated based on predefined patterns and cannot represent real-world program behaviors. Draper~\cite{draper}, Devign~\cite{devign} and CDG~\cite{vuldeepecker} were generated from real-world programs. However, as discussed in Sec.~\ref{sec:moti}, they suffer from labeling or source limitations. 
In fact, lacking good real-world datasets has become a major barrier for this field \cite{dl_vi_survey}. \sysname is automated and scales well on large real-world programs. It can produce more bug related information. We believe it can help to bridge the gap.

\smallskip

\noindent
\textbf{AI-based Static Analysis FP Reduction}. Static analysis is known to produce a lot of false positives. To suppress them,  several machine learning based approaches \cite{Kremenek_03, Jung_05, Yuksel_13, Hanam_14, ALETHEIA, Koc_17, Zhang_MAPL_17, Reynolds_17, Flynn_SEI, Raghothaman_18, Koc_19} have been proposed. Because they either target different languages or different static analyzers, they are not directly applicable. Inspired by their approaches, we designed and implemented a false positive reduction model for \infer as a use case for the \sysname dataset.



\section{Conclusion}
\label{sec:conclusion}

In this paper, we propose \sysname, a novel approach to label static analysis issues based on differential analysis and build a labeled dataset from real-world programs for AI-based vulnerability detection methods. We ran \sysname on 6 large programs and generated a labeled dataset of more than 1.3M examples with detailed bug related information obtained from the inter-procedural static analysis, the code base, and the commit history. By manually validating randomly selected samples, we show \sysname significantly improves the label quality compared to static analysis alone. We train a static analysis false positive reduction model as a use case for the \sysname dataset, which can effectively suppress false positives and help developers prioritize and investigate potential true positives first.

\bibliographystyle{IEEEtran}
\bibliography{paper}

\end{document}